\documentclass[hyper,12pt,letterpaper]{JHEP3}
\usepackage{epic,eepic}
\usepackage{amsmath,epsfig}
\usepackage[latin1]{inputenc}

\usepackage{latexsym}
\def\vev#1{\left\langle #1 \right\rangle}

\title{Degeneracy of Decadent Dyons}

\preprint{TIFR-TH-06-07}

\author{Atish Dabholkar$^{~1, ~2}$, K. Narayan$^{~3}$, and Suresh Nampuri$^{~1}$\\
{\it $^1$Department of Theoretical Physics}\\
{\it Tata Institute of Fundamental Research}\\
{\it Homi Bhabha Rd, Mumbai 400 005, India}\\

\it $^2${Laboratoire de Physique Th\'eorique et Hautes Energies (LPTHE)\\
\it{Unit\'e Mixte de Recherche (UMR 7589)}\\
\it{Universit\'e Pierre et Marie Curie-Paris 6; CNRS;\\}
\it{Tour 24-25, 5\` eme \'etage, Boite 126, 4 Place Jussieu, 75252
Paris Cedex 05}}\\

{\it $^3$Chennai Mathematical Institute}\\
{\it Padur PO, Siruseri 603103, \rm INDIA}\\}

\abstract{A quarter-BPS dyon in $\mathcal{N}=4$ super Yang-Mills theory is generically `decadent' in that it is stable only in some regions of the moduli space and decays on submanifolds in the moduli space. Using this fact, and from the degeneracy of the system close to the decay, a new derivation for the degeneracy of such dyons is given. The degeneracy obtained from these very simple physical considerations is in precise agreement with the results obtained from  index computations in all known cases. Similar considerations apply to dyons in $\CN=2$ gauge theories. The relation between the $\CN =4 $ field theory dyons and  those counted by the Igusa cusp form in toroidally compactified heterotic string is elucidated.}


\newcommand{\IR}{\mathbb{R}}
\newcommand{\IZ}{\mathbb{Z}}

\def\CN{{\cal N}}

\def\half{{\frac12}}

\def\IC{\relax\hbox{$\inbar\kern-.3em{\rm C}$}}

\def\IC{{\bf C}}

\def\CN{{\cal N}}

\def\bea{\begin{eqnarray}}
\def\eea{\end{eqnarray}}
\def\be{\begin{equation}}
\def\ee{\end{equation}}
\def\ba{\begin{align}}
\def\ea{\end{align}}
\def\bse{\begin{subequations}}
\def\ese{\end{subequations}}
\def\1F1{{}_1\!F_1}
\def\2F0{{}_2\!F_0}

\begin{document}

\section{Introduction}

In this note we consider the exact degeneracies of quarter-BPS dyons in $\mathcal{N}=4$ supersymmetric gauge theories. For a gauge group of rank $r$, the gauge group is broken to $U(1)^r$ on the Coulomb branch which is $6r$-dimensional for $\CN=4$. At a generic point in this Coulomb branch moduli space, there is a rich spectrum of such dyons in this theory whose degeneracy is known exactly in many cases from index computations and vanishing theorems as well as  from direct computations. Unlike the half-BPS dyons in $\mathcal{N} =4$ gauge theories which are stable in all regions of the moduli space, these dyons exist as stable single particle states only in some regions of moduli space. These dyons are prone to decay, or  are `decadent',  on  certain submanifolds of the moduli space, which can be of real codimension one or higher in $\CN=4$ theories.
We would like to know how `degenerate' these decadent dyons are.

The stability criterion for the decadent dyons follows from the usual considerations of charge and energy conservation using the BPS mass formula. For a dyon of electric charge vector $Q$ and magnetic charge vector $P$ we denote the total  charge vector by $\Gamma = [ Q;P]$. The BPS mass formula then gives the mass $M$ of such a state, in the $\CN =2$ notation, by the relation
\begin{equation}\label{bpsmass}
    M = |Z(\Gamma)| \, .
\end{equation}
where $Z(\Gamma)$ is the central charge that depends on the moduli fields and linearly on the charge vector $\Gamma$ . If the dyon with charge $\Gamma$ decays into two dyons with smaller charges $\Gamma_1$ and $\Gamma_2$ then one has $Z(\Gamma) = Z(\Gamma_1) + Z(\Gamma_2)$ which
by triangle inequality implies that
\begin{equation}\label{triangle}
    M = |Z(\Gamma)| \leq  |Z(\Gamma_1)| + |Z(\Gamma_2)| = M_1 + M_2.
\end{equation}
Hence, by energy conservation, the only way the decay can proceed is if
$M$ becomes equal to $M_1 + M_2$ at some point in the moduli space saturating the bound above. In $\CN=2$ theories, this  defines a codimension one surface or a `wall' in the moduli space. On one side of the wall where $M < M_1 + M_2$, the dyon with charge $\Gamma$ is stable. At the wall it is marginally unstable and decadent. Upon crossing the wall it no longer exists as a single particle stable state. In $\CN=4$ theories, there are more than one central charges which have to be aligned for the decay to occur and hence the submanifold of decadence can have codimension one or higher. As a result, this submanifold is not a wall since  one can just avoid it by going around it and access other regions of the moduli space. It is therefore more accurate it to call it the `surface of decadence' in the $\CN=4$ case.

Given such a dyon of charge $\Gamma$ that is stable in some region of the moduli space, we would like to know its degeneracy $\Omega(\Gamma)$ in that region. One can compute it  applying standard methods of semiclassical quantization of solitons in gauge theories, viewing the dyon as a charged excitation of a monopole system. Collective coordinate quantization then reduces the problem of computing the degeneracy $\Omega(\Gamma)$ to counting the number of eigenvalues of the  Hamiltonian of supersymmetric quantum mechanics of the bosonic and fermionic collective coordinates. This counting problem then becomes roughly equivalent to a cohomological problem of counting harmonic forms on the monopole moduli space which can be handled using index formulae\footnote{For quarter-BPS dyons, unlike in the half-BPS dyons, the problem is a little more subtle involving a potential on the moduli space as discussed in \cite{Stern:2000ie}.}.
Applying these methods, the degeneracy of quarter-BPS dyons
has been computed by Stern and Yi \cite{{Lee:1998nv}, {Bak:1999ip},{Bak:1999vd},Stern:2000ie} for a special
class of  charge assignments. The same formula has been derived from another  quiver dynamics in \cite{Denef:2002ru}.

We will give here a new derivation of the degeneracy of these dyons using a very simple physical argument that makes use of the fact that the dyons are decadent near the surface of decadence. We will utilize the known degeneracies of half-BPS dyons and an argument similar to the one used by Denef and Moore \cite{Denef:2007vg} in their discussion of the wall-crossing formula. The results are in perfect agreement with the known degeneracies of Stern-Yi dyons computed  using much more sophisticated techniques mentioned above. Moreover, this method can be naturally generalized to more complicated charge assignments as well as to  arbitrary gauge groups giving predictions for situations that have not hitherto been considered using the index methods.

The paper is organized as follows. In $\S{\ref{Computing}}$ we derive
the degeneracies of these dyons from their behavior near the surface of decadence. We present the basic physical argument in $\S{\ref{Basic}}$. We then discuss the case of $SU(3)$ Stern-Yi dyons in $\S{\ref{SU3}}$ and of $SU(N)$ Stern-Yi dyons in $\S{\ref{SUN}}$ and show that the degeneracy obtained using these arguments precisely agrees with the results known in these cases from index computations both in $\CN=4$ and $\CN=2$ cases. In $\S{\ref{String}}$ we discuss the relation of these dyons to the dyons in the field theory limit of string theory. We explain in particular why only some of the decadent dyons considered here are accounted for by the partition function for string theory dyons given by the inverse of the Igusa cusp form \cite{Dijkgraaf:1996it,{Gaiotto:2005gf}, {Shih:2005qf},{Gaiotto:2005hc}, David:2006yn, Sen:2007vb, {Dabholkar:2007vk}}. In $\S{\ref{Conclusions}}$ we conclude with comments.

\section{Computing degeneracies near the surface of decadence\label{Computing}}

For simplicity and also for comparison with known results, we consider in this section dyons  in $SU(N)$ gauge theories but these considerations are more  general and would apply to other groups.

It is well known that dyons in $SU(N)$ gauge theory have a nice geometric realization in terms of $(p, q)$ strings stretching between $N$ D3-branes.  The low energy world volume theory of $N$  D3-branes is a $U(N)$ Yang-Mills theory with $\mathcal{N} =4$ supersymmetry. Factoring out an overall center-of-mass $U(1)$ degree of freedom, one obtains an $SU(N)$ gauge theory. Simple roots of $SU(N)$ are $\{\alpha_i \}$ with $i = 1, \ldots N-1$ with the usual Cartan inner product
$ \alpha_i \cdot \alpha_i = 2 $, $ \alpha_i \cdot \alpha_j = -1$ for $i =  j \pm 1$, and $0$ otherwise.  Giving expectation values to the  six Higgs scalars in the adjoint representation corresponds to placing  the D-branes at non-coincident positions in the transverse $\mathbb{R}^6$ space which breaks the gauge symmetry to $U(1)^{N-1}$.

Consider a dyon with electric charge $Q$ and magnetic charge $P$ expanded in the basis of simple roots as
\begin{equation}\label{QP}
   Q = q_i \alpha_i, \quad P = p_i \alpha_i.
\end{equation}
If the electric and magnetic charge vectors are parallel to each other then the dyonic configuration preserves half the supersymmetries. Since it breaks eight supersymmetries, there are four complex fermionic zero modes for the center of mass motion giving rise to a 16-dimensional ultra-short multiplet. If the electric and magnetic charge vectors are nonparallel, the dyon preserves only a quarter of the supersymmetries. Since now it breaks twelve supersymmetries, there are six complex fermionic zero modes for the center of mass motion giving rise to a 64-dimensional short multiplet. In  $\mathcal{N}=2$ theories by contrast, in both cases, the dyon is half-BPS and there are four broken supersymmetries. Hence there are always two complex fermionic zero modes giving rise to a 4-dimensional half-hypermultiplet for the center of mass motion.

\subsection{Basic physical argument \label{Basic}}

Let us first summarize the argument for $\CN=2$ dyons of the type considered by Stern and Yi \cite{Stern:2000ie}. Given a dyon with charge $\Gamma = [ Q; P]$ of the Stern-Yi type, we would like to compute its degeneracy in a region of moduli space where it exists. Now, as we will discuss in the next sections, there exist surfaces of decadence for such a dyon where it decays into two dyons with charges $\Gamma_1 = [ {Q}_1 ; {P}_1]$ and $\Gamma_2 = [ {Q}_2 ; {P}_2] $ respectively.

Very close to the surface of decadence, the products of the decay are arbitrarily far away. In this case, one would expect that the degeneracy of the total configuration would be just the product of the degeneracies of individual fragments if the interactions between them were short-ranged.  However, this configuration has angular momentum in the long-ranged electromagnetic field
\begin{equation}\label{ang}
    J = \frac{1}{2} ( \vev{\Gamma_1 , \Gamma_2} -1),
\end{equation}
from the Saha effect as for a electron in the magnetic field of a magnetic monopole, where $\vev{\Gamma_1 , \Gamma_2} = Q_1 \cdot P_2 - Q_2 \cdot P_1$ is a symplectic product of charges that is invariant under $SL(2, \mathbb{Z})$ electric-magnetic duality. Note that there is a shift of $-1/2$ to the angular momentum of the electromagnetic field above, which has to do with the contribution of fermion zero modes \cite{Denef:2002ru}. Taking into account this additional degeneracy  of $(2 J+1)$ one concludes that the degeneracy of the original dyon is given by
\begin{equation}\label{degenN2}
    \Omega(\Gamma) =  |\vev{\Gamma_1 , \Gamma_2}| \, \Omega(\Gamma_1) \, \Omega(\Gamma_2).
\end{equation}
Note that the formula above counts the internal degeneracies, and hence does not include the overall multiplicity of four coming from the fermionic oscillators associated with the center of mass coordinate.
To get the total number of states, we must multiply (\ref{degenN2}) by this factor of $4$.

In the $\CN=4$ case there is an additional complication. In this case, we will be considering a decay in which one center, say $\Gamma_1$, is half-BPS. This center breaks eight supersymmetries. Since the overall state is quarter-BPS, the total configuration must break twelve supersymmetries. This can happen in two ways. Either, the center $\Gamma_2$ is quarter-BPS and breaks twelve supersymmetries by itself which includes the eight supersymmetries broken by the first center. Or, the center $\Gamma_2$ is half-BPS but breaks a different  half of the supersymmetries such that altogether there are twelve broken supersymmetries. In either case,  additional four supersymmetries are broken  in the internal theory of the two charge centers $\Gamma_1$ and  $\Gamma_2$. These broken supersymmetries give rise to two complex fermion zero modes that furnish a 4-dimensional multiplet with the same spin content as the half hypermultiplet of $\CN=2$. The degeneracy then is similar to  (\ref{degenN2}) with an additional multiplicative  factor of $4$:
\begin{equation}\label{degenN4S}
    \Omega(\Gamma) =  4 |\vev{\Gamma_1 , \Gamma_2}| \, \Omega(\Gamma_1) \, \Omega(\Gamma_2).
\end{equation}
To deduce this by a slightly different argument, one can think of the  total angular momentum of the system to be given by the tensor product of the half hypermultiplet with the spin $\bf j$ of the electromagnetic field given by (\ref{ang}). The half hypermultiplet has  spin content of one ({$\bf \half$}) + 2(\textbf{0}). The total system of the electromagnetic field and the relative zero modes has  spins $(\textbf{j + 1}) + 2 (\textbf{j}) + (\textbf{j - 1})$ with $\bf j$ given by (\ref{ang}). The multiplicity from these four representation is then $4 |\vev{\Gamma_1 , \Gamma_2}|$.

Let us now see how these formulae can be applied to compute the degeneracies of decadent dyons, for example, in the $\CN=4$ case. The formula (\ref{degenN4S}) effectively reduces the task of finding the degeneracy of $\Omega(\Gamma)$ of a state with charge $\Gamma$ to finding the degeneracies $\Omega(\Gamma_1)$ and $\Omega(\Gamma_2)$ of the subsystems. This in itself would not be  useful in general unless we knew how to compute $\Omega(\Gamma_1)$ and $\Omega(\Gamma_2)$ independently which is indeed the problem at hand. However, we will be considering the situation when at least one of the dyons with charge $\Gamma_1$ is  half-BPS and stable so that its electric and magnetic charges are parallel and are relatively prime. Such a dyon  we call irreducible, otherwise it is reducible.

Now, an irreducible dyon can be shown to have unit degeneracy using duality as follows. Since the electric and charge vectors are parallel,  we must have $Q_1 = a V_1$ and $P_1 = c V_1$ a primitive charge vector $V$. Further, since the dyon is an absolutely stable single particle half-BPS state, the integers $a$ and $c$ must be relatively prime for otherwise the dyon can split into subsystems without costing any energy.  Now, a primitive vector $V_1$  corresponds to a purely electric state and hence is proportional to the charge vector of a massive gauge boson of the theory. In this case, by an $SL(2, \mathbb{Z})$ electric-magnetic duality transformation, the state is dual to a purely electric gauge boson of the theory
\begin{equation}\label{primitive}
   \left(
     \begin{array}{cc}
       a & b \\
       c & d \\
     \end{array}
   \right)
    \left(
      \begin{array}{c}
        Q_1 \\
        P_1 \\
      \end{array}
    \right) =
    \left(
       \begin{array}{c}
         V_1 \\
         0 \\
       \end{array}
     \right).
\end{equation}
Since a  massive gauge boson of the theory is known to have unit degeneracy, by duality it then follows that  the half-BPS dyon with charge $\Gamma_1$ also has unit degeneracy. This conclusion can be explicitly checked also by a calculation similar to the one in  \cite{Sen:1994yi}.

The other decay product with charge $\Gamma_2$ can be either reducible or irreducible. If it is irreducible, then no further decay is possible. We then know the degeneracy of both decay products and hence of the original dyon using (\ref{degenN4S}). An example of such a decay when a quarter-BPS dyons goes directly into irreducible fragments will be discussed in $\S{\ref{SU3}}$ for $SU(3)$ dyons.

If the dyon with charge $\Gamma_2$ is reducible, then its  degeneracy is \textit{ a priori} not known. However, one can now apply
the reasoning in the previous paragraph iteratively. We can consider the surface of decadence of this dyon with charge $\Gamma_2$ where at least one of the decay products is irreducible. Continuing in this manner, one can relate the degeneracy of the original dyonic configurations to the degeneracies of the irreducible fragments up to factors coming from angular momentum degeneracies.  An example of such a decay will be discussed in the subsection $\S{\ref{SUN}}$ for $SU(N)$ dyons with $N>3$.

The reasoning outlined here is similar to the one used by Denef and Moore to derive the wall crossing formula for  dyons in $\CN=2$ string compactifications \cite{Denef:2007vg}. But there are differences. First, here we are using an additional input in the $\CN=2$ case  that on one side of the wall the degeneracy is zero. This can be ascertained for these field theory dyons from their realization as string webs. Second, for $\CN=4$ dyons, the surface of decadence is generically surface of codimension bigger than one and is not really a wall. So we are not  crossing any wall but merely approaching a surface of decadence. In $\CN=2$ string theories,  the dyon degeneracies are not known explicitly for a generic compactifications and there is no independent way of checking the validity of this reasoning. Here, in the context of supersymmetric gauge theories, explicit formulae are known for the degeneracies in the work of Stern and Yi. Our rederivation of the Stern-Yi degeneracies that we now describe in the following sections can thus be viewed as a check of the heuristic reasoning outlined above.

\subsection{Two-Centered Stern-Yi Dyons in $SU(3)$ Gauge Theories \label{SU3}}

Consider an  $SU(3)$ dyon in an $\mathcal{N}=4$ theory which has electric and magnetic charge vectors given by
\begin{eqnarray}
{ Q}&=& q_1 \alpha_1 + q_2 \alpha_2 \\
{P}&=& p_1 \alpha_1 + p_2 \alpha_2
\end{eqnarray}

Following earlier work of  \cite{Lee:1998nv} and \cite{Bak:1999ip}, Stern and Yi  \cite{Stern:2000ie} considered a simple charge configuration with magnetic charge vector ${P}= \alpha_1 + \alpha_2$. A quarter-BPS dyon can be viewed as a quantum charged excitation of a half-BPS monopole configuration.  Now if $q_1 = q_2$, then the electric and magnetic charge vectors of the dyon would be parallel, both along $\alpha_1 + \alpha_2$. Such a configuration would give a half-BPS state. To break the supersymmetry further and obtain a quarter-BPS dyon it is necessary that $s = q_1 -q_2$ is nonzero so that the electric and magnetic charge vectors are misaligned.
It is then useful to write the electric charge vector as
\begin{equation}\label{chargeQ}
    {Q} = (n+s)\alpha_1 + (n-s)\alpha_2.
\end{equation}
Dirac quantization condition then demands that $n \pm s$ must be integral although $n$ and $s$ could individually be half-integral\cite{Lee:1998nv}.
At some point in moduli space these states could decay into dyonic states into irreducible states
\begin{equation}
\left[(n+s) \alpha_1 + (n-s)\alpha_2 ; \alpha_1 + \alpha_2 \right]\rightarrow \left[ (n+s)\alpha_1; \alpha_1\right]+ \left[(n-s)\alpha_2; \alpha_2\right],
\end{equation}
so that $V_1 =\alpha_1$ and $V_2 = \alpha_2$ in the notation of the discussion in $\S{\ref{Basic}}$ and both decay products are irreducible.

Indeed in the string web picture \cite{{Schwarz:1996bh}, {Dasgupta:1997pu},{Sen:1997xi},{Aharony:1997bh},{Bergman:1997yw},Bergman:1998gs,Gauntlett:1999xz,{Argyres:2001pv}, Argyres:2000xs, Narayan:2007tx,Verlinde:2003pv}, the dyons are realized as a two-centered configuration. Near the surface of decadence the distance between the two centers becomes very large. Note that  the decay process across the wall is well described by semi-classical field configurations purely in terms of the low energy effective action on the Coulomb branch even when it occurs at strong coupling as would be the case for $\CN=2$ dyons \cite{Argyres:2001pv}.

Now since, both centers are half-BPS dyons, they  have unit degeneracy.  The contribution  from the angular momentum degeneracy factor is given by \begin{equation}\label{degen}
    |\vev{\Gamma_1 , \Gamma_2}| = |(n+s) \alpha_1 \cdot \alpha_2 - (n-s) \alpha_2 \cdot \alpha_1 |  = 2|s|
\end{equation}
Hence the degeneracy of a $SU(3)$ quarter-BPS dyon with charge vectors
$P = \alpha_1 + \alpha_2$ and $Q= (n+s)\alpha_1 + (n-s)\alpha_2$ is given by an application of the formula (\ref{degenN4S})
\begin{equation}
4 \cdot 2|s| \cdot 1 \cdot 1 =  8|s|,
\end{equation}
in precise agreement with the results of Stern and Yi. To get the total number of states, we multiply by a factor of $16$ coming from the center of mass multiplicity.

\subsection{Multi-centered Stern-Yi Dyons in $SU(N)$ Gauge Theory \label{SUN}}

We now consider more general  Stern-Yi dyons in a $SU(N)$ $\mathcal{N}=4$ gauge theory
where a cascade of decays is necessary to get to decay products that are all half-BPS. The charge vector is $\Gamma = \left[ Q; P\right]$ with
\begin{equation}
Q = (n + s_1 + \ldots  + s_{n-2})\alpha_1+(n-s_1 + \ldots + s_{n-2})\alpha_2 + \ldots (n- s_1 \ldots s_{n-2})\alpha_{n-1}\end{equation}
\begin{equation} P= \alpha_1 +\alpha_2 + \ldots + \alpha_{n-1}.\end{equation}
In the string web picture, these dyons are realized as multi-centered configurations.

We now approach the surface of decadence in the moduli space where the dyon  breaks up into half-BPS dyon with charge $\Gamma_1$ and a quarter-BPS dyon with charge $\Gamma_2$ given by
\begin{equation}\label{firstdecay}
    \Gamma_1 = \left[(n+ s_1 + s_2 + \ldots +s_{n-2})\alpha_1; \alpha_1 \right], \quad \Gamma_2 = \left[Q-Q_1; P-P_1\right].
\end{equation}
The angular momentum factor $\vev{\Gamma_1, \Gamma_2}$ equals $2s_1$. The $\Gamma_2$ charge center can further decay and we can iterate the process until we are left as the decay products with irreducible dyons of unit multiplicities. This iteration gives the degeneracy to be
\begin{eqnarray} \label{Stern}
16 \cdot  \prod_{i<j}^{N-2} |8s_i| \, ,
\end{eqnarray}
precisely what Stern and Yi obtained using their index computation.

For the $\CN=2$ dyons, similar reasoning using the formula (\ref{degenN2}) gives
\begin{eqnarray} \label{Stern2}
4 \cdot  \prod_{i<j}^{N-2} |2s_i|,
\end{eqnarray}
once again in agreement with Stern and Yi.

\section{Relation to string theory dyons \label{String}}

The partition function that counts the degeneracies of quarter-BPS dyons in heterotic string theory compactified on a six-torus ${\bf T}^6$ is given in terms of the Igusa cusp form which is a modular form  of weight ten of the group $Sp(2, \mathbb{Z})$. It depends on  three complex variables with a Fourier expansion given by
\begin{equation}\label{igusa}
   \frac{1}{\Phi_{10}(p, q, y) } = \sum c(m, n, l) p^m q^n y^l,
\end{equation}
where the sum is over $ m, n \geq -1$  and $l \in \mathbb{Z}$.
A quarter-BPS dyons in this theory is specified by  a charge vector $\Gamma = (Q_e; Q_m)$ where here both $Q_e$ and $Q_m$ are Lorentzian
vectors that take values in the $\Gamma^{22, 6}$ Narain lattice. There are three quadratic combinations   $Q_e^2, Q_m^2, Q_e \cdot Q_m$ with respect to a Lorentzian inner product invariant under the $O(22, 6,; \mathbb{Z})$.
For a given vector $Q_e$ in this lattice, one can define
the right-moving part $Q_{eR}$ to be the projection onto the $22$ space-like directions and $Q_{eL}$ to be the projection onto the $6$ time-like directions. The inner product is then defined by
\begin{equation}\label{inner}
    Q_e^2  = Q_{eR}^2 - Q_{eL}^2.
\end{equation}
The degeneracy $d(\Gamma)$ is then given  in terms of the Fourier coefficients by
\begin{equation}\label{degen2}
    d(\Gamma) = c(Q_e^2/2, Q_m^2/2, Q_e \cdot Q_m).
\end{equation}

This formula was proposed in \cite{Verlinde:1997fm} and derived in \cite{Shih:2005qf,David:2006yn} using the 4d-5d lift and using a genus-two partition function in \cite{Gaiotto:2005hc}. Generalization to CHL orbifolds have been discussed in \cite{{Jatkar:2005bh},{Dabholkar:2006bj},{Dabholkar:2006xa}, {David:2006ud},{David:2006ru},David:2006ji}. Note that according to the prescription above, we can have nonzero degeneracies apparently only for states that have
\begin{equation}\label{bound}
Q_e^2 \geq -2, \quad Q_m^2 \geq -2.
\end{equation}
A more careful treatment of the degeneracy formula extends them by analytic continuation to all other states related by electric-magnetic duality to those that satisfy $Q_e^2 \geq -2$ and $Q_m^2 \geq -2$ in a way that the spectrum is duality invariant \cite{Dabholkar:2007vk, {Sen:2007vb},{Cheng:2007ch}}.

Since the low energy effective action for the heterotic string contains the action for supersymmetric nonabelian Yang-Mills theory, it is  natural  to ask if the dyon partition function above also counts degeneracies of these decadent dyons that we have considered in the previous sections. Indeed, our work was partly motivated by this question. If this is true, it would give a nontrivial check of the degeneracies predicted by the dyon partition function.

If the dyon partition function could count the field theory dyons like the Stern-Yi dyons then it would lead to many puzzles. Firstly, 
the degeneracies derived from the dyon partition function depend only
on the three integers $Q_e^2, Q_m^2, Q_e \cdot Q_m$ and not on the components of the charges as the Stern-Yi degeneracy (\ref{Stern}) seems to depend on. Second, the Stern-Yi degeneracies only grow polynomially
as a function of charges, whereas the stringy dyon degeneracy grows exponentially if the discriminant
\begin{equation}\label{discri}
    \Delta = Q_e^2 Q_m^2 - (Q_e \cdot Q_m)^2,
\end{equation}
is positive.

We will show that these puzzles get resolved by the fact that the field theory dyons are in a different duality orbit than the ones that are counted by the dyon partition function. Hence one cannot apply the dyon partition function to count the Stern-Yi dyons except for special ones when the gauge group is $SU(3)$.

To see this clearly, we need to think more carefully about the field theory limit of string theory. To be able to analyze a dyon in field theory we would like to decouple stringy states and gravity from the consideration. At the same time, we would like to have a nonabelian structure in the gauge theory so that we do not have to deal with a Dirac monopole which is a singular field configuration but have instead a t'Hooft-Polykov monopole. In this case, the monopole is smooth solitonic configuration with a finite core which can be analyzed in field theory using semiclassical quantization. Such a limit is easily achieved if we consider the gauge group like $SU(3)$ to be embedded in the left-moving $E_8 \times E_8$ symmetry for example and consider Higgs expectation value  $v$ that is much smaller compared to  the string mass scale $\Lambda$. In this case massive string states can be ignored.  Moreover, the mass $M$ of dyons will go as $v/g^2$ where $g$ is the string coupling and gravitation backreaction will go as $GM^2 = v^2/\Lambda^2$ using the fact Newton's constant $G$ goes as $g^2/\Lambda^2$. Thus, gravitational back reaction can also be ignored as long as $v$ is much smaller than $\Lambda$ and one can analyze the dyons in a field theory limit.

It is crucial for a useful field theory limit that the charges are purely left-moving, that is $Q_e^2 < 0$ and $Q_m^2 <0$. This is because, in the heterotic string, which consists of a right-moving superstring and a left-moving bosonic string, only the left-moving $U(1)$ gauge symmetries can get enhance at special points in the moduli space of toroidal compactification. For example, for a circle compactification, at a generic radius of the circle we have $U(1)_L \times U(1)_R$ which couples to the charges
\begin{equation}\label{charges}
    q_{L,R} = \sqrt{\frac{\alpha'}{2}} ( \frac{m}{R} \pm \alpha' w R),
\end{equation}
where $m$ is the Kaluza-Klein momentum and $w$ the winding number along the circle. At the self-dual radius of the circle however where $R^2 = \alpha'$, only $U(1)_L$ gets enhanced to a nonabelian $SU(2)_L$ but the $U(1)_R$ remains abelian. This is a consequence of the fact that the left-moving ground state energy is $-1$ as the bosonic string whereas the right-moving ground state energy is $0$ as for the superstring. As a result, while certain states carrying left-moving momentum become massless at the self-dual radius, all states carrying right-moving momentum remain massive.

This implies that dyons coupling to both  right-moving and left-moving $U(1)$ fields cannot be analyzed in a field theory limit as nonsingular solitonic configuration and stringy corrections would have to taken into account. For this reason we should embed our field theory gauge group into the purely left-moving symmetry.

Following, this reasoning, we can  embed an $SU(N)$ gauge group  into the $SO(32)$ gauge group of the heterotic string for $N \leq 16$. In this case $Q_e = Q$ and $Q_m = P$ with $Q_e^2 = -Q^2$ and $Q_m^2 = -P^2$ with the understanding that the $Q_e^2$ and $Q_m^2$ are defined using the Lorentzian inner product (\ref{inner}) whereas $Q^2$ and $P^2$ are defined using the positive definite Euclidean Cartan metric on the root space of the gauge group as we have used in the previous sections \footnote{Once we turn on Wilson lines to break the gauge group we will have more accurately $Q_e = Q + k$ where $k$ is a light-like vector with $Q \cdot k =0$ so that $Q_e^2$ still equals $-Q^2$. Moreover, the charge vector is not strictly left-moving. This does not change the main point of the argument and hence we will ignore it.}. We refer to the charge vector as spacelike, timelike, or lightlike depending on whether the Lorentzian norm is positive, negative, or zero respectively. With the embedding above, we conclude that the field theory dyons must correspond to states with timelike charge vectors in the Narain lattice.

To understand the main issues, let us first focus on the $SU(3)$ Stern-Yi dyons. For the degeneracy of a string theory dyon that satisfies the bound(\ref{bound}) to match with a Stern-Yi dyon, the two charge configurations must lie in the same U-duality orbit. Now, the U-duality group $G(\IZ)$ of the string theory is
\begin{equation}\label{uduality}
    G(\IZ) = O(22, 6,; \mathbb{Z}) \times SL(2, \IZ).
\end{equation}
The U-duality orbit of the charges can be characterized by various  invariants. To start with, we have the discriminant defined in (\ref{discri}) which is the unique quartic invariant of the continuous duality group $G(\IR)$. In addition, as  noted in \cite{Dabholkar:2007vk}, there is a discrete invariant
\begin{equation}\label{discrte}
  I =  gcd(Q_e \wedge Q_m).
\end{equation}
The wedge product gives the antisymmetric area tensor of the parallelogram bounded by the vectors $Q_e$ and $Q_m$. The invariant
$I$ then counts the number of lattice points inside this parallelogram
\cite{Dabholkar:2007vk}. See also \cite{{Banerjee:2007sr},{Banerjee:2008ri}}.

For an $SU(3)$ Stern-Yi dyon with charge vectors are
\begin{equation}\label{chargeQP}
   P = \alpha_1 + \alpha_2, \quad  {Q} = (n+s)\alpha_1 + (n-s)\alpha_2.
\end{equation}
Using the embedding described above, we see that the two invariants for such an configuration are given by
\begin{equation}\label{syinivariant}
    \Delta = 12 s^2, \quad I = 2s
\end{equation}

We note that $\Delta>0$ as it must be for a BPS configuration. Now, starting from spacelike $Q_e^2$ and $Q_m^2$, one can show that its impossible to go by U-duality to a configuration with both electric and magnetic charges timelike. To prove this we consider a general S-duality transformation acting on $Q_e$ and $Q_m$ as
\begin{eqnarray} Q'_e &=& a Q_e + b Q_m\\Q'_m &=& c Q_e + d Q_m \end{eqnarray} Now, if $Q_e$ and $Q_m$ are positive norm vectors then $aQ_e \pm bQ_m$ is a positive norm vector. So, ${Q'_e}^2 \geq 0$ and similarly ${Q'_m}^2 \geq 0$. Thus, the only string dyonic configurations which can be U-dual to a field theory dyon will be those with timelike $Q_e$ and $Q_m$\footnote{The other possibility is having the electric(magnetic) charge to be timelike and the magnetic(electric) charge to be spacelike. But this will yield $\Delta<0 $ and breaks supersymmetry.}. By definition of $\Phi_{10}$, the only such charges it counts are those with $Q_e^2 = -2$ and  $Q_m^2 = -2$. Taking $Q_e \cdot Q_m = M$ to be arbitrary, we obtain a dyonic charge configuration with invariants $I=1$ and $\Delta = 4- M^2$.
Hence the two sets of invariants match only for $s={1\over 2}$ and $M=\pm 1$ which corresponds to  $Q_e^2=-2$, $Q_m^2 =-2$ and $|Q_e \cdot Q_m|=1$. Consequently, only these string dyonic configurations lie in the duality orbit of SU(3) Stern-Yi dyons.

It is easy to see that in fact \textit{all} Stern-Yi dyons with electric charge  {\ref{chargeQ} are counted by the dyon degeneracy formula for all values $n$ with $s =1/2$.  This follows from the fact that one can change the value of $n$ by a duality transformation of the form
\begin{equation}\label{tduality}
    \left(
      \begin{array}{cc}
        1 & n - \half \\
        0 & 1 \\
      \end{array}
    \right)
\end{equation}
Note that  $n$ must be half-integral for configuration with $s =1/2$.

For an $SU(N)$ Stern-Yi dyonic configuration with $N>3$ given by \begin{eqnarray}Q_e &=& \sum^{N-1}_{i=1} (n+\sum^{N-2}_{j=1}P_{ij}s_j)\alpha_i\\Q_m &=& \sum^{N-1}_{i=1}\alpha_i\end{eqnarray} where $P_{ij}= -1$ for $j<i$ and $P_{ij}=1$ for $i\geq j$, the U-duality invariants are $\Delta=4(2 \sum^{N-2}_{i=1}{s_i}^2 +\sum^{N-2}_{i,j=1} s_i s_j)$ and $I = gcd(2s_1,2s_2,..2s_{N-2})$. For matching to the configurations whose degeneracy is counted by $\Phi_{10}$ we must have $I=1$ which translates to the condition  that the various $2s_i$ are mutually coprime. Further, we can easily see that $\Delta >3$ for these Stern-Yi dyons\footnote{This follows easily from the inequality $({s_i}^2+{s_j}^2) > 2 s_i s_j$.}. Hence the string theory dyons do not lie in the U-duality orbit of any field theory SU(N) dyon with $N>3$.

We therefore conclude that with the exception of the SU(3) dyons with $I=1$, the field theoretic dyons considered earlier are outside the realm of applicability of the dyon partition function of string theory dyons in terms of the Igusa cusp form. A similar analysis has been carried out independently in \cite{{Sen:2007ri},Banerjee:2007ub}. For other values of $I>1$, a different partition function is required. For a recent proposal for the dyons with I=2 see \cite{Banerjee:2008pv}.

\section{Conclusions \label{Conclusions}}

We have seen that a simple physical argument allows one to compute the degeneracies of decadent  dyons in $\CN=2$ and
$\mathcal{N}=4$ supersymmetric Yang-Mills theory with little work.
These results are in agreement with the known results obtained using much more elaborate and sophisticated index computations. Our results could also be viewed as a test of the reasoning underlying the wall-crossing formula in $\mathcal{N}=2$ theories and of the degeneracy formula near the curve of decay in $\mathcal{N}=4$ theories. This method of course allows one to count decadent dyons with more general charges in general gauge groups not hitherto considered in the field theory literature. It would be interesting to test such predictions using  index computations.

It may seem surprising that this almost classical computation is capable
of capturing the quantum degeneracies precisely. In this context, we note that a number of essentially quantum ingredients have implicitly gone into our reasoning. First, the shift of $-1/2$ to the classical field angular momentum from the fermionic zero modes in \ref{ang} is essentially quantum. Second, the angular momentum multiplicities of $2J+1$ are also quantum. What is interesting is that after incorporating this information into an almost classical reasoning, one can determine the degeneracies exactly.

Finally, we have also seen that the dyons counted in field theory are not accounted for by the dyon partition functions recently derived in the context of string theory dyons except for one special case. This is because they lie in a different duality orbit than the dyons for which the dyon partition function has been derived.

\subsection*{Acknowledgements}

It is a pleasure to thank Jeff Harvey, Kimyong Lee,  Anindya Mukherjee, Sameer Murthy, Rahul Nigam, Suvrat Raju, Ashoke Sen, and Piljin Yi for valuable discussions.

\bibliographystyle{JHEP}
\bibliography{decadent}

\end{document}